\documentclass[twocolumn]{aastex7}
\usepackage{amsmath}
\shorttitle{The shape and velocity of black hole coronas}
\shortauthors{Krawczynski, Hu, and Chun}
\begin{document}
\title{The Shape and Velocity of Black Hole Coronas}

\author[orcid=0000-0002-1084-6507,gname='Henric',sname='Krawczynski']{Henric Krawczynski}
\affiliation{Washington University in St. Louis, the McDonnell Center for the Space Sciences and the Center for Quantum Leaps, St. Louis, MO 63130}
\email[show]{krawcz@wustl.edu}  

\author[orcid=0000-0003-0350-3132,gname='Kun',sname='Hu']{Kun Hu}
\affiliation{Washington University in St. Louis, the McDonnell Center for the Space Sciences and the Center for Quantum Leaps, St. Louis, MO 63130}
\email[show]{hkun@wustl.edu}  

\author[orcid=0009-0002-2488-5272,gname='Sohee',sname='Chun']{Sohee Chun}
\affiliation{Washington University in St. Louis, the McDonnell Center for the Space Sciences and the Center for Quantum Leaps, St. Louis, MO 63130}
\email[show]{sohee.chun@wustl.edu}  

\begin{abstract}
The first highly significant detection of the polarization of the X-rays from the Black Hole X-ray Binary Cygnus X-1 revealed surprisingly strong polarization parallel to its radio jet.
The observation is now commonly interpreted as resulting from the Comptonization of some lower frequency seed photons in a laterally extended corona. The emitting plasma is likely 
moving along the jet axis with a sizable fraction $\beta$ of the speed of light, amplifying the polarization degree.
In this letter we address the question of how laterally extended the corona actually needs to be to account for the observations, and we calculate the required outflow velocities of velocity-stratified coronas as a function of the stationary-frame aspect ratio $h=H/2R$ of  cylindrical coronas of height $H$ and radius $R$. We find that models with $h\le0.8$ and $\beta_{\rm top}\,=\,0.5-0.9$ can give both the observed photon index of $\Gamma\,=\,1.6$ and the observed 4\% polarization parallel to the symmetry axis. 
We do not find suitable solutions for larger $h$.
Given the constraints on $h$, we discuss the likely 
size of the corona, emphasizing that compact coronas 
close to the black hole are preferred as they 
can tap more efficiently into the gravitational 
energy and the rotational energy of the black hole, 
and are less exposed to their own emission reprocessed in the optically thick portion of the accretion flow.
\end{abstract}
\section{Introduction}
The Imaging X-ray Polarimetry Explorer {\it IXPE} \citep{2024Galax..12...38W,2024mbhe.confE..31C} measured the hard state polarization of Cyg X-1 to be (4.01$\pm$0.20)\% over the 2-8 keV band parallel to its radio jet \citep{2022Sci...378..650K}. 
The polarization of the accreting black hole may even be somewhat higher as the reflection of the X-rays off the stellar companion and its wind is likely to reduce the observed net polarization \citep{vandermeulen2026predictions}.
While the Comptonization of longer wavelength 
seed photons in a laterally extended corona 
explain the polarization angle (PA), 
the polarization degree (PD) predicted at the
system inclination of 27.51$^{\circ}$ is only 1-2.5\% 
\citep{2022Sci...378..650K,2023ApJ...949L..10P}, 
short of the observational result. 
Following the detection of the optical 
polarization parallel to the radio jets of AGNs, \citet{1998ApJ...496L.105B} identified a 
mechanism to explain the high PDs more than 20 years prior to the launch of {\it IXPE}: the Comptonization in a plasma outflowing 
with mildly relativistic velocities 
(in units of $c$) of $\beta\sim0.5$ \citep[see also][]{2001MNRAS.326..417M}. 
Following the {\it IXPE} observations, the model was invoked for the hard state by 
\citet{2023ApJ...949L..10P} and for soft state 
by \citet{2025ApJ...993...54K,2026A&A...711A.281N}. 
Alternative explanations of the high PD include: 
a higher inclination of the inner accretion flow and thus the Comptonizing region than the system inclination \citep{2022Sci...378..650K}; the Comptonization by plasmoids in current sheets \citep{2026ApJ..1008L..37G}; and the reflection of the X-rays off a thicker outer disk \citep{2023ApJ...957....9W} or an accretion disk wind \citep{2024MNRAS.527.7047T}, or both \citep{2026MNRAS.547ag498T}.     

In this letter, we explore the outflowing corona model more in depth. It is attractive owing to its conceptual simplicity as it does not require the addition of another poorly defined accretion flow component, i.e., a wind or a thick outer disk. Radiation pressure \citep{1999ApJ...510L.123B}, 
magnetic buoyancy \citep{1979ApJ...229..318G},
magnetic tension and pressure, as well as magneto-centrifugal forces \citep{2007A&ARv..15....1D} can all accelerate pairs, and to a lesser degree electron-ion plasma, to relativistic velocities. General relativistic magnetohydrodynamic (GRMHD) simulations of accreting black holes reveal another mechanism: magnetic energy accumulates close to the event horizon and causes violent ejections that accelerate plasma to high velocities; in addition, magnetic reconnection generates 
fast moving plasmoids or flux ropes \citep{2024MNRAS.527.6065G,2025ARA&A..63..127S}.

The corona amplifies the seed photon power by the factor of $A$, and the energy balance between heating and cooling the corona then dictates the photon number spectral index $\Gamma$  \citep[][called HM in the following]{1991ApJ...380L..51H,1993ApJ...413..507H}. 
For some corona geometries, i.e., slab coronas, 
radiative feedback limits the produced energy spectra to be softer than the observed ones \citep[HM;][]{2026A&A...706A..78D}.
Relativistic motion of the corona with velocity $\beta$ 
away from the equatorial plane of the black hole suppresses 
the feedback, and thus alleviates the problem.
The feedback reduces to the fraction $f_\beta\propto(1-\beta)^2$, with the exact value of $f_\beta$ depending on the relative locations of the corona and the reflecting 
and reprocessing portion of the accretion flow 
\citep[see Equations (1)-(5) of][]{1999ApJ...510L.123B}. 

The actual coronas of Black Hole X-ray Binaries (BHXRBs) and Active Galactic Nuclei (AGNs) are almost certainly located close to the black hole, and thus do not have an indefinitely large lateral extension (finite $h$ instead of $h\rightarrow0$).
We thus performed a series of simulation runs to determine which corona aspect ratios $h$ are compatible with 
the existing spectral and polarimetric data. 
We will describe our model in Section \ref{s-methods}, 
and discuss the results in Section \ref{s-h-results}.

Our aim is to take the observational data to constrain not only the shape but also the size and location of the corona.
As mentioned above, the corona is likely energized close 
to the black hole to tap into its gravitational and 
rotational energy \citep{2025ApJ...993...54K}. 
A compact corona close to the black hole is 
attractive as they lead to steeply falling emissivity profiles
matching the observed ones 
\citep{2015ApJ...813...84G,2015ApJ...808....9P,2021ARA&A..59..117R}, and capture less reflected and reprocessed emission.
They thus maximize the emission of broad emission 
lines and minimize the radiative feedback problem.
See related discussions from the pre-{\it IXPE} era in \citep[][and references therein]{2020A&A...644A.132U,2020MNRAS.498.3302W}.
We review the results and discuss implications for the location and size of the corona in Section \ref{s-discussion}.

\section{Radiation transport calculations}
\label{s-methods}
We model coronas with different aspect ratios with an extended version of the {\tt polrad} Monte Carlo polarized 
radiation transport code \citep{2025ApJ...993...54K}. 
The extended code, called {\tt polrad-3D}, transports photons in the stationary 
frame through a 3-D cylinder with its symmetry axis 
aligned with the $z$-axis. The code implements 
different velocity profiles along the $z$-axis.
We initially neglect general relativistic effects as the 
location and the size of the corona as well as the properties of the accretion flow are still uncertain.
We will discuss the effect of the general relativistic 
effects on the inferred corona parameters in 
Section~\ref{s-discussion}. See \citep[][]{2026arXiv260424961Z,Chun2026} for discussions 
focusing on specific scenarios and geometries. 
While {\tt polrad-3D} is a simpler version of 
{\tt kerrC} which accounts for curved spacetime
\citep[][]{2022ApJ...934....4K}, the results 
are more general in the sense that they are 
not limited to a specific location and 
extent of the corona.

We assume an outflowing corona with radius $R$ and height $H\,=\,2\,h\,R$ with a stationary location and  
shape. The continuity equation then implies a stationary frame density $n(z)\,\propto\,\beta^{-1}$. 
Denoting the average density of the electrons or pairs of the corona as $\bar{n}$, the vertical and horizontal depths of the corona are $t_z\,=\,\bar{n}\sigma_{\rm T}H$ and 
$t_r\,=\,\bar{n}\sigma_{\rm T}R$ 
with $\sigma_{\rm T}$ being the Thomson cross sections.
We study constant coronal outflow velocities $\beta=\beta_0$ 
as well as two models with the flow accelerating 
from $\beta_0\,=\,0.1$ to $\beta_{\rm top}$. 
The first model is motivated by the radiation 
pressure driven velocity profiles of \citet{1999MNRAS.305..181B} 
and was also adopted by \citet{2023ApJ...949L..10P}. 
It assumes $\beta\propto\tau$ except for an 
initial layer where $\beta\,=\beta_0$ 
where the plasma is launched.
The velocity of the plasma is given by:
\begin{eqnarray}
\beta(s)& = &\max\left[\beta_0, \beta_{\rm max}\sqrt{s}\right]
\end{eqnarray}
with $s$ being the fractional coronal height $s\,=\,z/H$.
$\beta(s)$ is constant for 
$s<s_0$ with $s_0\,=\,(\beta_0/\beta_{\rm max})^2$.
We can derive the integrated depth $\tau$ as a function of 
$s$ (or $z$) and the velocity as a function of $\tau$ and obtain:
\begin{align}
N &\equiv \int_{0}^{1}\frac{ds}{\beta(s)} \,=\, \frac{2\beta_{\mathrm{max}}-\beta_0}{\beta_{\mathrm{max}}^{\,2}} \\
\tau(s) &= 
\begin{cases}
\frac{t_z s}{N\beta_0} & s<s_0 \\
\frac{t_z}{N}\left[\frac{2\sqrt{s}}{\beta_{\mathrm{max}}}-\frac{\beta_0}{\beta_{\mathrm{max}}^{2}}\right] & s\ge s_0\\
\end{cases}\\
\tau_{0} &= \frac{\beta_0}{2\beta_{\mathrm{max}}-\beta_0}\,t_z\\
\beta(\tau) &=
\begin{cases}
\beta_0, & \tau < \tau_{0}\\
\dfrac{\beta_0}{2} + \left(\beta_{\mathrm{max}} - \dfrac{\beta_0}{2}\right)\dfrac{\tau}{t_z}, & \tau \ge \tau_{0}
\end{cases}
\end{align}
We find that the results are rather insensitive to the choice of $\beta_0$ and we use $\beta_0\,=\,0.1$.
Note that $\tau_0$ is 
the depth at $s_0$.

We use a second model in which $\beta$ is linear in $s$:
\begin{eqnarray}
\beta(s)&=& \max{\left[\beta_0,\beta_{\max}s\right]}. 
\end{eqnarray}
The main difference from the first to the second profile is that the first accelerates faster; the second has a larger launch layer, and then picks up velocity quicker.
We assume the electrons or pairs follow a Maxwell J\"uttner distribution of a plasma at temperature 100\,keV. 

To keep the assumptions of the model as simple as possible, we assume that the seed photons are cyclo-synchrotron photons generated by the coronal electrons and inject them with an energy of 10 eV 
in the plasma frame. 
The 10\,eV is the synchrotron self-absorption turnover: for 
a $k_{\rm B}T\, =\, 100$\,keV plasma ($k_{\rm B}$ is the Boltzmann constant) in a $B\, \approx\, 10^7$\,G magnetic field (the field in equipartition with the coronal radiation for $R \approx r_{\rm g}$, see Section \ref{s-discussion}), the emission is self-absorbed below 10\,eV.
The non-absorbed photons are emitted by $\gamma\,=\,10$ electrons at around 10\,eV
\citep{2000MNRAS.314..183W,2009MNRAS.392..570M,2013MNRAS.430.3196V}.
We use a rejection technique to implement a constant emissivity per electron or pair  in the fluid frame making use of the fact that the lab frame emission rate scales as $\propto n/\gamma$ with  $\gamma=(1-\beta^2)^{-1/2}$. 
We launch unpolarized photons isotropically in the plasma frame.
The photons' wave four vectors $k^\mu$, polarization four vectors $f^\mu$ are subsequently transformed from the plasma rest frame into the stationary frame with a Lorentz boost.
The polarization degree is a Lorentz invariant.  

The photon transport is implemented using the 
infinitesimal invariant optical depth \citep{2025ApJ...993...54K}. 
Using asterisks for the plasma frame quantities, we have 
$d\tau\,=\,n^*\sigma_{\rm T}\,dl^*$.
Note that $d\tau$ is not directly related to $\tau$ above, 
which merely gives the amount of plasma in the corona, and 
$dl^*$ is a line element along the photon path and 
not related to $s$ above. 
The line elements $dl^*$ in the plasma frame 
and that in the stationary frame $ds$ are related by
$dl^*\,=\,\gamma(1-\beta\mu)\,dl$ with $\mu$ being the 
cosine of the stationary frame angle between the plasma and
photon directions. We thus get:
\begin{equation}
    d\tau\,=\,n(z)\,\sigma_{\rm T}\,(1-\beta(z)\mu)\,dl
\end{equation}
Photons are transported from the current location in the corona $s_1$ to the next $s_2$  by generating the depth to the 
next collision  $\tau_c\,=\,-\ln u$ from a uniform random number  $u\in\left(0,1\right]$. We use a pre-calculated table of 
\begin{equation}
\tau(s_1\rightarrow s_2)\,=\,\int_l \frac{n}{\bar{n}}(1-\beta\mu)\,dl'
\end{equation}
to evaluate how far a photon propagates in $s$ by solving $\tau(s_1\rightarrow s_2)\,=\,\tau_c$ for $s_2$.
Photons escaping the cylinder through the top or sides are used to derive the observational signatures. We assume that photons escaping through the bottom of the cylinder 
are lost to the black hole. Evaluating the impact of the radiative feedback from coronal photons reflecting and reprocessing in the optically thick parts of the flow is outside of the scope of this paper.  
If the photon remains within the cylinder we transport 
it by a total distance of $l\,=\,H(s_2-s_1)/\mu$ 
into its current direction, and scatter it.
The scattering engine \citep{2017ApJ...850...14B} 
transforms $k^\mu$ and $f^\mu$  first into the reference 
frame of the outflow, then into the reference frame of the scattering electron. After scattering the photon with Fano's
scattering matrix (expressed in terms of Stokes parameters) that accounts for the full Klein  Nishina cross section 
\citep{1961RvMP...33....8M}, 
we perform the inverse transformations 
back into the plasma frame and back into the stationary frame.  
We sort the events exiting through the top or the sides of the cylinder into inclination bins and store their Stokes $I$, $Q$ and $U$ parameters. The Stokes parameters are referenced to a coordinate system in which the $Q>0$ direction is aligned with the black hole and accretion disk angular momentum vectors.
Owing to the assumed axial symmetry of the corona, $U\,=\,0$, and the PD equals $q\,=\,Q/I$ with positive values
denoting the electric field vector polarization direction parallel to the symmetry axis. 
Results from validating the code versus the literature are shown in Appendix \ref{a:verification}.  

We initially run the simulations over a coarse grid in $t_z$ and $\beta_{\rm top}$ with $3.2\times 10^5 $ photons per bin, and measure the 2–30 keV photon index $\Gamma$ as well as the 2-8\,keV $q$-value. 
We subsequently interpolate the results to estimate the $t_z$ giving $\Gamma=1.6$, and evaluate a dedicated run at the estimated $t_z$.
The result is accepted if $|\Gamma - 1.6| \le 0.02$ and correct the residuals based on $d\Gamma/d \ln t_z$. Along the $\Gamma=1.6$ contour, we then search for the smallest $\beta_{\rm top}$ giving $q\,=\,$4\%.
Runs with 6.4$\times 10^8$ simulated photons were used to obtain the final result.  
\section{Results - limits on the corona's aspect ratio}
\label{s-h-results}
Our first objective is to find the parameter configurations that give a valid description of two key observational results of 
Cyg X-1 during the 2022 {\it IXPE}, {\it NICER} and {\it NuSTAR} observation campaign \citep{2022Sci...378..650K}: a 2-30\,keV photon index of $\Gamma\,=\,1.6$ and a 2-8\,keV polarization degree of PD\,$=\,$4.01\%$\pm$0.2\%. The fits satisfying the 2-8\,keV {\it IXPE} polarization constraints automatically also fit the  15-60\,keV XL-Calibur polarization results which have 
much larger error bars \citep{2025ApJ...994...37A}.
We thus ran the code over a grid of starting parameters 
to find for each $h\in \left\{0, 0.1, 0.2, 0.4, 0.8, 1.6, 3.2, 6.4, 12.8\right\}$ which $t_z$ and $\beta_{\rm top}$ combination gives the observed PD and $\Gamma$. 

\begin{figure*}
  \centering
  \includegraphics[width=0.95\textwidth]{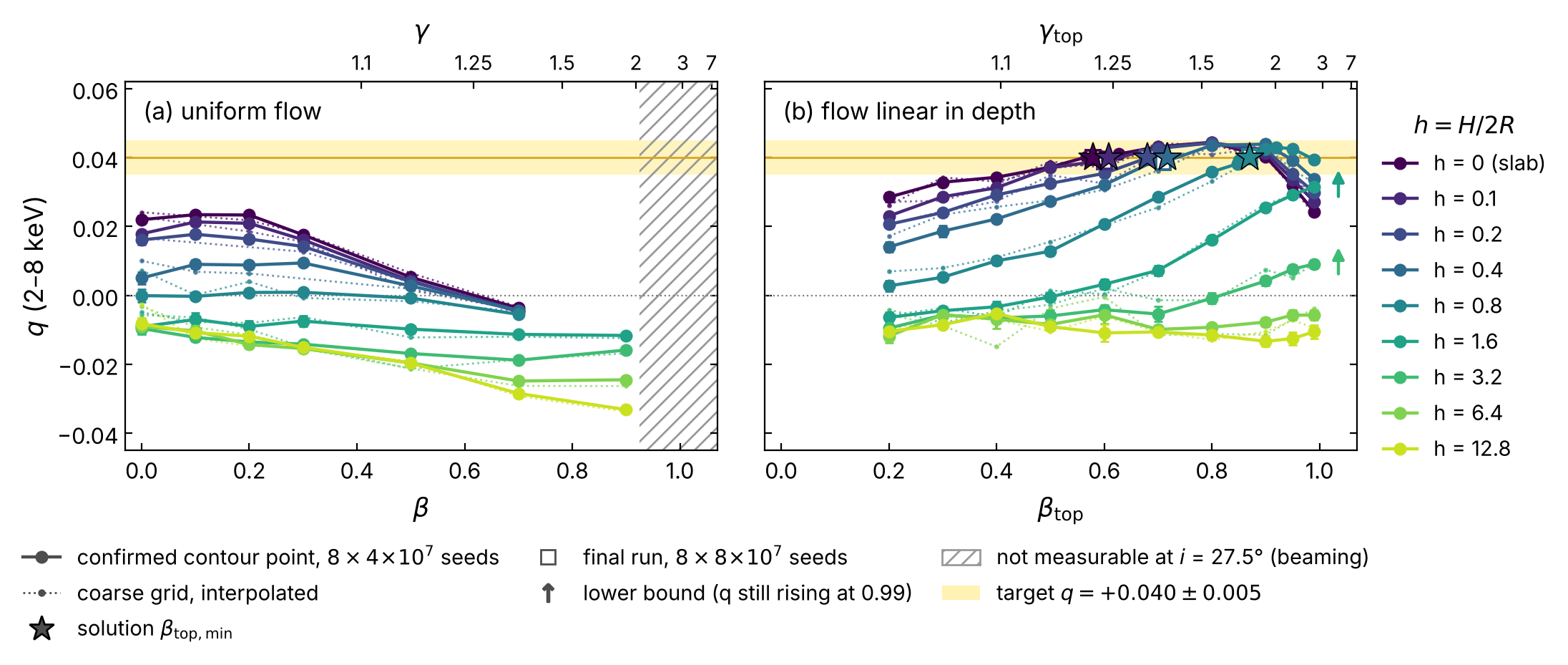}
  \caption{Polarization of constant velocity coronas (left panel) and coronas with the plasma accelerating from $\beta_0\,=\,0.1$ to $\beta_{\rm top}$ linearly in depth (right panel). The polarization is given in terms of the normalized Stokes parameter $q=Q/I$, positive (negative) for the electric field polarization angle parallel (perpendicular) to the $z$-axis. The dotted lines show results from interpolating the results from the original calculation grid. The solid lines and data points are from the final simulations. All results show models which give $\Gamma=1.6$.}
  \label{f:q1}
\end{figure*}

Figure \ref{f:q1} shows how the polarization at the inclination of Cyg X-1 for different aspect ratios $h$ as a function of the velocity $\beta$ for the constant velocity corona (left panel), and for coronas with plasma accelerating linearly with depth $\tau$ (right panel).  Both panels show the results of the models giving a photon index of $\Gamma\,=1.6$.
The left panel shows that the polarization tends to decrease rather than to increase with $\beta$ for the constant-$\beta$ models. We identified the origin of the decrease by analyzing the $q$-values at matched aberrated inclinations. For coronas with small $h$, aberration alone would increase $q$. 
While maintaining $\Gamma\,=\,1.6$ fixes the optical depth in the direction of the observer, the optical depth in all other directions increases with $\beta$, giving a weaker polarization.
For $h\,\ge\,0.8$ the co-moving view of the emitting column corona is more strongly polarized perpendicular to the axis than parallel to it. The right panel shows that the coronas with accelerating plasma behave fundamentally different and the polarization $q$ increases, at least initially, 
with $\beta_{\rm top}$. 
In this case, the faster electrons or pairs see the photons emitted by slower portions of the plasma. The scattering of photons coming in their reference frame from the top of the corona, give the large polarization degrees parallel to the $z$-axis. We see that the maximum $q$ that we can get can flip sign (see $h=1.6,3.2$) for sufficiently fast upper corona flows. 
For the highest $\beta_{\rm top}$-values, $q$ decreases again.
In these models, the emission of the fastest layer is so boosted that it misses the observer at $i\,=\,27.51^{\circ}$.
The de-boosted emission from the fastest layers and the emission from the slower layers gives lower $q$-values.   
The results for the velocity linear in height rather than in depth look similar to the one linear in depth.

\begin{figure*}
  \centering
  \includegraphics[width=0.65\textwidth]{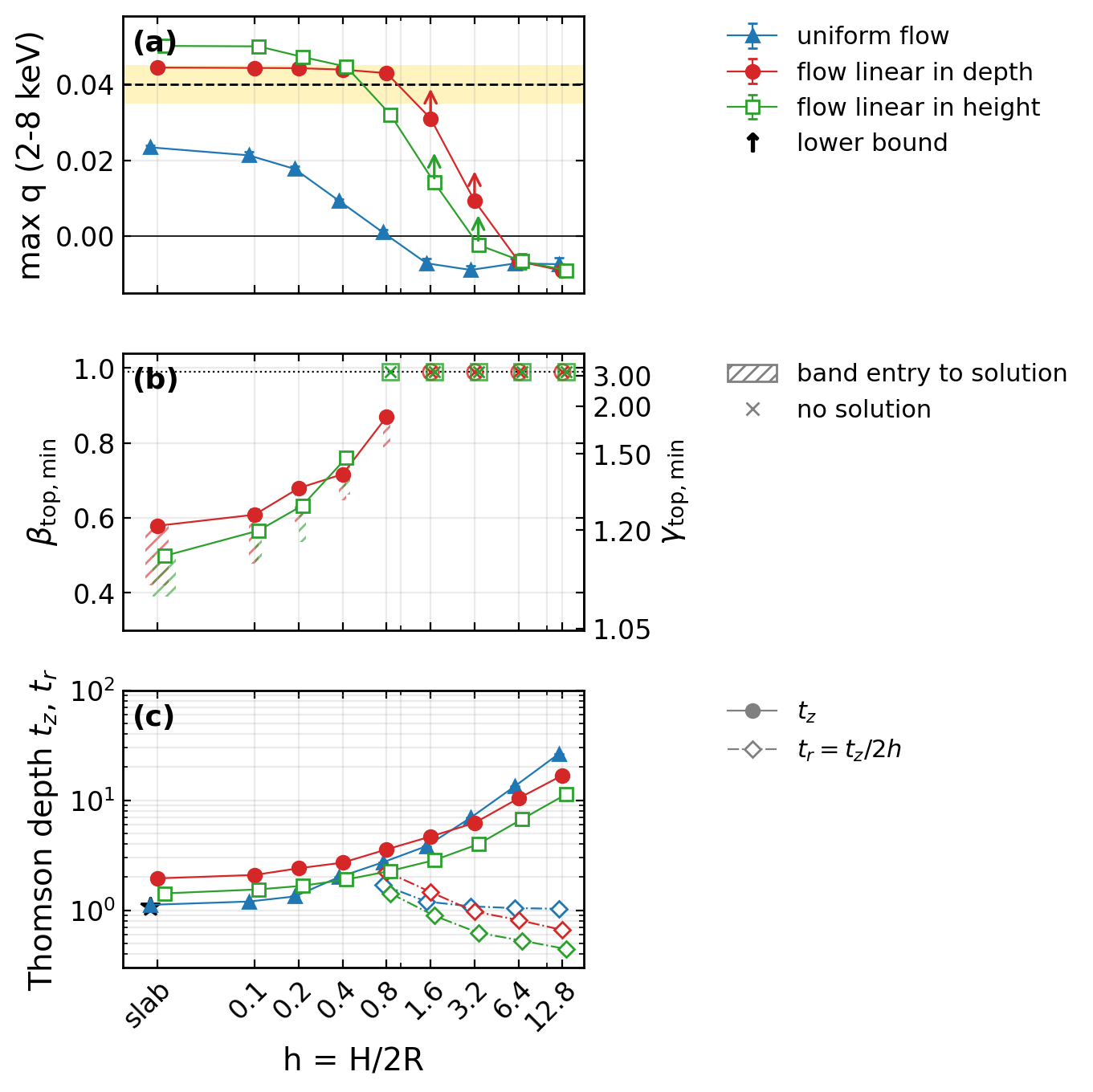}
  \caption{The top panel shows the highest polarization $q$ on the $\Gamma=1.6$ contour
  at a viewing inclination of 27.5$^{\circ}$ 
  as a function of the aspect ratio of the 
  corona in the stationary frame for outflows 
  with constant $\beta$ (blue lines), 
  $\beta$ linear in depth (red lines), 
  and linear in height (green lines). 
  For aspect ratios $h\le0.8$, we find solutions matching the observed $q=4$\%. 
  The yellow band shows the observed value 
  with a 0.5\% systematic uncertainty band.      
  The center and lower panels are only for the velocity stratified coronas. The center panel shows the $\beta_{\rm top}$ required to match the observed $q$ for $h\le0.8$, as well as the $\beta_{\rm top}$-values giving the highest $q$ for h$\ge$1.6. The hatched regions show the values entering the yellow band of 4\%$\pm$0.5\%.   
  The right $y$-axis translates the $\beta_{\rm top}$-values into the corresponding Lorentz factors. The lower panel shows the vertical (and for $h\ge 0.8$ the radial) depths of the models achieving $q\,=\,$4\%, or, for models that fail, the depths that give the  highest PD while $\Gamma\,=\,1.6$.}
  \label{f:q2}
\end{figure*}

Figure \ref{f:q2} summarizes the most important results. The top panel shows the maximum $q$-values achieved with the three model families (constant velocity, velocity linear in depth, and velocity linear in height) for the models with the 2-30\,keV photon index of $\Gamma\,=\,1.6$.
We see that models with $h\,\le 0.8$ can reproduce or even exceed the observed $q$-value of 4\%.
For $h\ge1.6$ we do not find models that match the data and we give the maximum $q$-values at the edges of our sampled parameter space.  The center panel shows the velocity at the top of the corona $\beta_{\rm top}$ that increases from $\sim$0.5 to 0.9 for $h$ rising from 0 to $0.8$. 
For larger $h$-values, we plot the upper edge of the parameter space that we explored ($\beta=0.99$) as lower limits. 
The lower panel shows the vertical depth $t_z$ for the 
models that explain the observations for $h\le0.8$, and the depths at the highest $q$ on the $\Gamma=1.6$ contour
for $h\ge1.6$. For models with $h>1.6$ we also show the average radial depth $t_r$. We see that $\tau$ increases as $h$ and thus  $\beta_{\rm top}$ rise to compensate 
for the decrease of the optical depth 
owing to the alignment of the electrons and photons in the flow.

\begin{table*}
\footnotesize
\caption{Solutions ($q=+0.040$ on the $\Gamma=1.6$ contour, $i=27.5^{\circ}$) and maxima if there is no solution. For each velocity profile we give $\beta_{\rm top,min}$ and $\gamma_{\rm top,min}$ if a solution exists, otherwise the highest $q$ on the contour and the $\beta_{\rm top}$ at which it occurs (in brackets), and the vertical Thomson depth $t_z$ at that point, with $t_r = t_z/2h$ in brackets for $h \ge 0.8$. ``lb'' (lower bound): $q$ still rising at $\beta_{\rm top}=0.99$.}
\label{tab:sol}
\hspace*{-2cm}
\begin{tabular}{@{}lllllll@{}}
\hline\hline
$h$ & $\beta$ linear in depth & $t_z$ ($t_r$) & $\beta$ linear in height & $t_z$ ($t_r$) & uniform flow, $q_{\max}$ & $t_z$ ($t_r$) \\ \hline
0    & \parbox[t]{3.6cm}{\raggedright $\beta_{\rm top,min}=0.579\pm0.011$, $\gamma_{\rm top,min}=1.23$} & 1.95 & \parbox[t]{3.3cm}{\raggedright $\beta_{\rm top,min}=0.499\pm0.006$, $\gamma_{\rm top,min}=1.15$} & 1.42 & \parbox[t]{2.6cm}{\raggedright $+0.0234\pm0.0004$ ($\beta=0.1$)} & 1.12 \\
0.1  & $0.609\pm0.014$, 1.26 & 2.09 & $0.566\pm0.008$, 1.21 & 1.55 & \parbox[t]{2.6cm}{\raggedright $+0.0213\pm0.0009$ ($\beta=0.1$)} & 1.20 \\
0.2  & $0.680\pm0.004$, 1.36 & 2.41 & $0.633\pm0.016$, 1.29 & 1.67 & \parbox[t]{2.6cm}{\raggedright $+0.0177\pm0.0008$ ($\beta=0.1$)} & 1.33 \\
0.4  & $0.717\pm0.011$, 1.43 & 2.71 & $0.761\pm0.008$, 1.54 & 1.90 & \parbox[t]{2.6cm}{\raggedright $+0.0094\pm0.0004$ ($\beta=0.3$)} & 2.00 \\
0.8  & \parbox[t]{3.6cm}{\raggedright $0.870\pm0.007$, 2.03 (marginal)} & \parbox[t]{1.9cm}{\raggedright $3.58\pm0.14$ ($2.23\pm0.09$)} & \parbox[t]{3.3cm}{\raggedright none: $q_{\max}=+0.0320\pm0.0008$ ($\beta_{\rm top}=0.9$)} & 2.28 (1.42) & \parbox[t]{2.6cm}{\raggedright $+0.0010\pm0.0007$ ($\beta=0.3$)} & 2.72 (1.70) \\
1.6  & \parbox[t]{3.6cm}{\raggedright none: $q_{\max}=+0.0310\pm0.0007$ ($\beta_{\rm top}=0.99$, lb)} & 4.68 (1.46) & \parbox[t]{3.3cm}{\raggedright none: $+0.0141\pm0.0011$ ($\beta_{\rm top}=0.99$, lb)} & 2.85 (0.89) & \parbox[t]{2.6cm}{\raggedright $-0.0071\pm0.0013$ ($\beta=0.1$)} & 3.82 (1.19) \\
3.2  & \parbox[t]{3.6cm}{\raggedright none: $q_{\max}=+0.0093\pm0.0010$ ($\beta_{\rm top}=0.99$, lb)} & 6.22 (0.97) & \parbox[t]{3.3cm}{\raggedright none: $-0.0023\pm0.0009$ ($\beta_{\rm top}=0.99$, lb)} & 3.99 (0.62) & \parbox[t]{2.6cm}{\raggedright $-0.0089\pm0.0011$ ($\beta=0$)} & 6.93 (1.08) \\
6.4  & \parbox[t]{3.6cm}{\raggedright $q<0$ everywhere; least negative $-0.0066\pm0.0012$ ($\beta_{\rm top}=0.3$)} & 10.4 (0.81) & \parbox[t]{3.3cm}{\raggedright $q<0$ everywhere; $-0.0066\pm0.0021$ ($\beta_{\rm top}=0.8$)} & 6.76 (0.53) & \parbox[t]{2.6cm}{\raggedright $-0.0072\pm0.0013$ ($\beta=0$)} & 13.4 (1.05) \\
12.8 & \parbox[t]{3.6cm}{\raggedright $q<0$ everywhere; least negative $-0.0091\pm0.0012$ ($\beta_{\rm top}=0.4$)} & 16.9 (0.66) & \parbox[t]{3.3cm}{\raggedright $q<0$ everywhere; $-0.0090\pm0.0011$ ($\beta_{\rm top}=0.9$)} & 11.35 (0.44) & \parbox[t]{2.6cm}{\raggedright $-0.0074\pm0.0016$ ($\beta=0$)} & 26.3 (1.03) \\ \hline
\end{tabular}
\end{table*}
Table \ref{tab:sol} gives the values of the solutions (or limits) shown in Figure \ref{f:q2}.

We studied the dependence of our results on our assumptions for our favored model ($\beta$ increasing linearly with depth). The results do not depend strongly on the velocity threshold $\beta_0$. We find approximately the same $q$-results for $\beta_0\,=\,0.05$ and $\beta_0\,=\,0.1$ on the $\Gamma\,=\,1.6$-contour. Changing the absorbing lower surface to a surface reflecting 50\% of the impinging photons reduces the optical depths of the $\Gamma\,=1.6$ contour, but not the $q$-values on the contour. 
\section{Discussion}
\label{s-discussion}
Our study shows that coronas with aspect ratios $h\le$0.8 and the coronal plasma accelerating from 0.1$c$ to 0.5-0.9$c$ 
can describe the hard energy spectrum  ($\Gamma\,=\,$1.6) 
and strong polarization ($q\,=\,$4\%) of Cyg X-1 in the hard state. The corona models discussed in this letter produce 
their own synchrotron seeds. They emit X-rays strongly polarized parallel to the black hole and disk angular 
momentum vectors owing to the velocity stratification 
with faster layers scattering the photons produced 
by slower layers. The net outcome is similar to a directed outflow scattering seed photons isotropic in the stationary reference frame as discussed in \citep{2024MNRAS.528L.157D}. 

In the following we discuss the possible location and physical properties of such a corona. We  assume the Cyg X-1 system parameters 
with $M=21.2\,M_\odot$ and $r_{\rm g}=3.1\times10^6$~cm \citep{2021Sci...371.1046M}. 
The Eddington Luminosity is $L_{\rm Edd}\,=\,4\pi\,G\, M\, m_{\rm p}\,c\,/\,\sigma_{\rm T}\,=\,2.67\times10^{39}$~erg~s$^{-1}$.
We assume that the corona has an aspect ratio of $0.4$ and 
the observed photons scatter last in layers moving at $\beta\,=\,0.5$.
The Doppler factor of photons from the corona reaching the observer at inclination $i$ is: $\delta\,=\,\left[\gamma(1-\beta \cos{i})\right]^{-1}\approx 1.56$.
For a time averaged hard state luminosity of 
$L_{\rm iso}\,=\,0.005L_{\rm Edd}$ \citep{2002ApJ...578..357Z,2003MNRAS.343L..84G}, the
corona frame luminosity is:
\begin{equation}
    L\,=\,\frac{\gamma}{\delta^3}L_{\rm iso}\,\approx\,0.3\times L_{\rm iso}
\end{equation}
The coupling of the plasma and the radiation and the pair creation and annihilation are dictated by the compactness parameter of the corona with radius $R$:
\begin{equation}
\ell\equiv
\frac{L\sigma_{\rm T}}{R\,m_{\rm e}\,c^3}
=4\pi\frac{m_{\rm p}}{m_{\rm e}}\frac{L}{L_{\rm Edd}}\frac{r_{\rm g}}{R}
=
35.4 \frac{L_{\rm iso}}{0.5\%\,L_{\rm Edd}}(r_g/R)
\label{eq:ell}
\end{equation}
Pair creation inside the corona \citep{1984MNRAS.209..175S,1995ApJ...449L..13S} 
depends on $R$ and a  number of additional factors.
As discussed below, the compactness may be larger than estimated here, and the coronal temperature may be 
larger than the observed 100\,keV cutoff of the 
Comptonized emission if the corona is located close to the black hole so that the gravitational redshift has to be taken into account. The observed non-thermal power law tail 
may play a rather minor role in creating pairs if 
it cuts off at a few MeV \citep{2002ApJ...572..984M}. 
The recent study of \citet{2026arXiv261105307R} indicates that reconnection can accelerate particles so that they emit a hard synchrotron component extending to the 
magnetic-field-independent 16 MeV synchrotron burnoff energy (in the local stationary frame). Such a non-thermal tail 
can easily load the corona with pairs. 
The upcoming COSI mission \citep{2022hxga.book...73T} 
can test if the MeV 
tail is strongly polarized as predicted in this scenario
\citep{2026arXiv261105307R}. 

The Compton cooling time $t_{\rm IC}\,=\,R/c\ell$ 
is 0.03$(R/r_{\rm g})R/c$, i.e., shorter than the light crossing time, 
indicating that the particles are heated as they radiate. 
The synchrotron boiler mechanism starts with a power law distribution of electrons or pairs which build up a thermal distribution through interacting with their
own cyclo-synchrotron emission 
\citep{2009MNRAS.392..570M}.
Alternatively, turbulent energy can cascade 
and produce a mix of random kinetic bulk motion as well as microscopically heated plasma 
\citep{2024PhRvL.132h5202G,2026ApJ..1001...64G}.
If the magnetic field provides the energy to heat the corona, we can estimate the magnetic field from the assumptions of equipartition between the radiation field and the magnetic field. With the magnetic compactness:
\begin{equation}
\ell_B \equiv \frac{\sigma_{\rm T}\,R\,u_B}{m_{\rm e}c^2}
      = \frac{\sigma_{\rm T}\,R\,B^2}{8\pi\,m_{\rm e}c^2}
      = 10.1\,\left(\frac{B}{10^{7}\,{\rm G}}\right)^{\!2}\frac{R}{r_{\rm g}}\,,
\label{eq:ellB}
\end{equation}
equipartition between the radiation and magnetic field energy densities $u_{\rm rad}\,=\,u_{\rm B}$ gives $\ell_{\rm B}\approx \ell/3$ and 
\begin{equation}
 B_{\rm eq}\,\approx\,1.1\times\,10^7\,(R/r_{\rm g})^{-1}{\,\rm G\,}.
\label{e:b}
\end{equation}
This magnetic field transports a Poynting flux of $L_B\,=\,(B^2/4\pi)\,\gamma^2\,\beta c\,\pi R^2\,\approx\,0.43\,L_{\rm iso}$ independent of $R$ - sufficiently large to power the corona.
In the scenario of a magnetically powered pair corona, the magnetic field carries the bulk of the energy flux. 
The energy flux carried by pairs is $L_{e^+e^-}\,=\,\gamma^2\,\beta\,c\,n^*\,m_{\rm e}\,c^2\,\pi\,R^2$. The magnetization is thus $\sigma\,=\,L_B/L_{e^+e^-}\,=\,B^2/(4\pi n^* m_{\rm e}c^2)$. With $n^*\,=\,t_r/\sigma_{\rm T}R$, we get $\sigma\,=\,2\ell_B/t_r\,\approx\,6.9\,(R/r_{\rm g})^{-1}$, 
showing that $L_B\gg L_{e^+e^-}$. For this magnetization,
the Alfv\'en speed and thus the plasmoid bulk velocity is 
$v_{\rm A}\approx \sqrt{\sigma/(\sigma+1)}\,\approx\,0.94\,c$ \citep{2017ApJ...850..141B,2024MNRAS.527.6065G}, 
consistent with the coronal velocities in our model. 
Including the kinetic energy of the 100\,keV plasma 
in the calculation gives a 55\% higher $L_{e^+e^-}$, 
a hot magnetization smaller by the same factor, and an Alfv\'en
velocity of 0.9$c$. The magnetically dominated scenario only works for a pair corona. The energy flux of an ion-electron 
corona would be larger by a factor $m_{\rm p}/m_{\rm e}$, suppressing $\sigma$ to values $\ll$1.  

In magnetically powered corona models, Faraday rotation limits the pair content of the corona \citep[see also][]{
2009ApJ...703..569D,2021ApJ...914...51K,
2024ApJ...977..201B,2025A&A...701A.115K,
2026ApJ...998L..47K,
2026A&A...705A...8M}. After their last scattering, photons cross on average one Thomson free path $1/n\sigma_{\rm T}$, leading to a Faraday rotation by:
\begin{equation}
\begin{split}
\Delta\chi&=\frac{e^3\lambda^2g}{2\pi m_{\rm e}^2c^4\sigma_{\rm T}}\,B_\parallel\,\eta_{\Delta}\\
&\simeq10\,\frac{B_\parallel}{10^7\,{\rm G}}\left(\frac{\rm keV}{E}\right)^{\!2}\eta_\Delta\ {\rm rad},
\end{split}
\end{equation}
where $\lambda=\delta\lambda_{\rm obs}$ is the wavelength 
in the plasma frame. The factor $g=K_0(1/\Theta)/K_2(1/\Theta)=0.70$ is a relativistic 
correction at 100~keV \citep{2008ApJ...688..695S}, and the $K$'s denote the modified Bessel functions of the second kind, and $\Theta\,=\,k_{\rm B}T/m_{\rm e}c^2$.
The factor $\eta_{\Delta}$ is the asymmetry of electrons and positrons:
\begin{equation}
\eta_{\Delta}\,=\,\frac{n_--n_+}{n_-+n_+},
\end{equation}
with $\eta_{\Delta}\approx 1$ for an electron-ion plasma
and $\eta_{\Delta}\ll 1$ for a pair plasma.
The polarization angles of the {\it IXPE} observation averaged over the 2-3\,keV and the 3-4\,keV bins are $-17.1^{\circ}\pm2.2^{\circ}$ and $-20.4^{\circ}\pm2.3^{\circ}$,
respectively \citep[Figure S5 of][]{2022Sci...378..650K}; 
the difference of $3.3^{\circ}\pm3.2^{\circ}$ is consistent 
with a constant angle at the $\approx 1\,\sigma$ level.
Requiring the Faraday rotation to differ by less than $3.3^{\circ}+3.2^{\circ}=6.5^{\circ}=0.11$\,rad
between the two bins ($\Delta (E^{-2})\,=\,0.082$\,keV$^{-2}$) gives
$\eta_\Delta\,<\,0.14\,(B_{\parallel}/10^{7}\,{\rm G})^{-1}$ 
at the one-sigma confidence level. 
The result excludes an electron-ion plasma ($\eta_{\Delta}\,=\,1$) unless the magnetic field is by a factor of $\ge 8$ below the
equipartition field of Equation (\ref{e:b}), and constrains the purity of the pair plasma: for the equipartition field at $R\,=\,r_{\rm g}$, $\eta_{\Delta}\,<\,0.13$, 
i.e., fewer than one ionization electron per $\approx8$ 
leptons in the layer of the last scatterings.
The argument gets weaker for tangled magnetic fields.

Above we assumed that $R \sim r_{\rm g}$ as it seems unlikely that the flow can concentrate the required energy into much smaller regions. We assume throughout this letter that the corona 
feeds itself with synchrotron photons. As we have shown here, the scenario leads to a self-consistent magnetically-fed 
corona of size $R\sim r_{\rm g}$.

An alternative scenario involves a larger corona intercepting just the right amount of the soft accretion disk flux to explain the observed Comptonized flux while still allowing for $\Gamma\,=\,1.6$. We can estimate the coronal size at which disk seed photons might become important. The Comptonization with $\Gamma\,=\,1.6$ from the seed photon energy of $E_s\sim 0.5$\,keV (3 times the color temperature of $T_{\rm d}\,=\,0.17$\,keV of the disk) 
to $E_c\sim 100$\,keV corresponds to a seed power 
amplification of $A\,\approx\, (E_c/E_s)^{2-\Gamma} =(100{\rm\,keV}/0.5 {\rm\,keV})^{0.4}\,\approx\,8$
\citep[close to the results of][]{1999ApJ...510L.123B}.
We infer $L_s\,=\,L/A\approx 0.036\, L_{\rm iso}\,\approx\,5\times10^{35}{\rm \, erg\,s^{-1}}$.
The disk emits the power $\pi R^2\, \sigma_{\rm SB}\, T_{\rm d}^4$ towards the corona, with $\sigma_{\rm SB}$ being the Stefan--Boltzmann constant. 
The static corona with moving plasma intercepts this flux. 
Transforming the luminosity into the plasma frame gives  
$dL^{*}\,=\,\gamma^2(1-\beta\mu)^2\,dL$ \citep[][]{1999ApJ...510L.123B}. 
We evaluate the seed photon luminosity in the frame of the plasma moving with $\beta\,=\,0.5$ in which the observed photons scatter last, i.e., in the same 
frame used for computing $L$ and $\delta$. 
Averaging over the Lambertian disk intensity gives
$D_{\rm s}\,=\,\gamma^{2}\left(1-\frac{4}{3}\beta+\frac{1}{2}\beta^{2}\right)\,=\,0.6$.
The first scatterings occur in slower layers where  $D_{\rm s}\,=\,0.7-0.9$. The relative Lorentz factor between the different layers is small however, and the calculation using
a single $\beta$ is accurate to $\approx\!15\%$. 
Solving for $R$:
\begin{eqnarray}
R&\gtrsim&\left(\frac{L/A}{D_{\rm s}\,\pi\sigma_{\rm SB}T_{\rm d}^{4}}\right)^{\!1/2}\\
&=&5.6\,r_{\rm g}\!\!
\left(\frac{L}{0.3\,L_{\rm iso}}\right)^{\!\!1/2}\!\!\!
\left(\frac{A}{8}\right)^{\!\!-\frac{1}{2}}\!\!\!
\left(\frac{D_{\rm s}}{0.6}\right)^{\!\!-\frac{1}{2}}\!\!\!
\left(\frac{k_{\rm B}T_{\rm d}}{0.17\,{\rm keV}}\right)^{\!\!-2}\!.
\label{eq:seedradius}
\end{eqnarray}
Note that the static velocity stratified corona is easier to feed with seed photons than a corona made of distinct ejections, as these would see a lower seed photon flux.
The required radius of the corona is larger if the corona covers the disk at larger $r$ where the disk 
temperature is lower, or the disk is truncated.   

The discussion so far neglected general relativistic effects. 
If the corona is compact and located close 
to the black hole these effects are not entirely negligible.  
For a stationary corona positioned at $r\,=\,r_{\rm C}\,r_{\rm g}$ on the symmetry axis of the black hole with spin parameter $a$, 
the local time $\tau$ and the coordinate time $t$ are related to each other via:
\begin{equation}
g\,\equiv\,\frac{d\tau}{dt}\,=\,\sqrt{-g_{00}}\,=\,\sqrt{\frac{\Delta}{r_{\rm C}^2+a^2}}
\end{equation}
with $g_{00}$ the 0-0 element of the metric in Boyer-Lindquist coordinates, and $\Delta\,=\,r_{\rm C}^{2}-2\,r_{\rm C}+a^{2}$ 
\citep{1967JMP.....8..265B,1972ApJ...178..347B}.
Photons emitted along the $z$-direction arrive at an observer at infinity with the energy lowered by the same factor $g$.  
For $a\,=\,0.98$ and $r_{\rm C}\,=\,5$ and 10 we get $g\,=\,0.78$ and 0.90, respectively. The local luminosity and the compactness of Equation~(\ref{eq:ell}) both scale with $g^{-2}\,=\,1.6$ and 1.25 for $r_{\rm C}\,=\,5$ and 10, respectively. 
The equipartition field of Equation~(\ref{e:b}) and the co-moving temperature scale with $g^{-1}$ and thus rise
by 28\% and by 11\% for $r_{\rm C}\,=\,5$ and 10, respectively. Pair creation in the corona depends more strongly on $g$ as two effects combine: $\ell\propto g^{-2}$ and the local temperature inferred from the measured temperature scales as $\propto g^{-1}$.
A corona close to the black hole is thus more likely to feed itself with pairs. In terms of the polarization of the coronal emission, the impact of gravitational lensing rapidly 
declines as $r_{\rm C}$ increases from 5 to 10.
More precise estimates require specific corona models combined 
with ray tracing simulations as in \citep[][]{Chun2026,2026arXiv260424961Z}.  Overall, we can conclude that, while general relativistic effects are not negligible, they do not change the conclusions qualitatively.

As mentioned in the introduction, the moving corona beams its emission in forward direction, and therefore reduces the radiative feedback from emission reflected by and reprocessed by the disk. The flip side of the reduced radiative feedback has weaker Fe $K\alpha$ emission lines than 
without the beaming. A logical next step to the work presented in this paper would be to model not only $\Gamma\,=\,1.6$ and $q\,=\,4$\%, as we have 
done in this paper, but to calculate 
the energy spectrum of the reflected emission with the aim to reproduce the simultaneously 
measured Fe K$\alpha$ line profile as well.  
In this context, we would like to note that the broad 
Fe K$\alpha$ line does not necessarily need to be produced 
close to the black hole. \citet{2026ApJ..1005..119N} model the line with a strongly $r$-dependent ionization profile and find that most of the emission originates rather far ($r\ge 10\,r_{\rm g}$) from the black hole.

In summary, a corona with aspect ratios $h\,=\, H/2R\,\le\,0.8$, with a radius of $R\sim r_{\rm g}$ and with plasma accelerating from 0.1$c$ to $(0.5-0.9)c$ can explain the hard energy spectrum ($\Gamma\,=\,1.6$) and the strong polarization ($q\,=\,$4\%) of Cyg X-1 in the hard state rather naturally. 

This letter focuses on BHXRBs. Note that most corona properties are scale invariant \citep[see for example the equations of][]{2008ApJ...688..555G}. The main difference between BHXRBs and AGNs is the mean energy of the disk and coronal seed photons. They impact the amplification factor $A$, and the energies at which coronal gamma-rays pair produce. Another difference is the much weaker Faraday rotation in AGN coronas \citep{2026ApJ...998L..47K}. The main results of this paper, i.e., the constraints on the aspect ratio and velocities of viable corona models carry over to AGNs. 

\section*{Use of AI in this paper:}
The project was conceived and planned out by the authors, and the entire text of the paper was written by the authors. 
We used Claude Fable 5 (model ID claude-fable-5-1) to help extend {\tt polrad}  to {\tt polrad-3D} and to discuss the results of the study. 
We used an agentic pipeline similar to \citep{2026arXiv261014185D} to run the code over a matrix of starting values, to
iterate the solutions until sufficiently precise solutions were found, and to write code to 
plot the results.       
All code and results were vetted by the 
authors and they take full
responsibility for the results.

\section*{Data Availability}
The {\it polrad-3D} code and all the {\tt Python} used for the figures in this paper will be made available through a public git and through a Zenodo record as soon as the paper is accepted for publication.    
\begin{acknowledgments}
The authors thank Yajie Yuan, Alex Chen, John Groger,
Shravan Vengalil Menon, and Chris Done for joint discussions. 
H.K.\ acknowledges NASA support from the grants 80NSSC24K1178, 80NSSC24K1749, and 80NSSC24K1819, as well as support from the 
Washington University in St. Louis Physics Department in the framework of the Hufferd seed funding program, and the McDonnell Center for the Space Sciences at Washington University in St. Louis.
\end{acknowledgments}

\appendix
\section{Verification of the {\tt polrad-3D}}
\label{a:verification}

\begin{figure*}[b]
  \centering
  \includegraphics[width=0.85\textwidth]{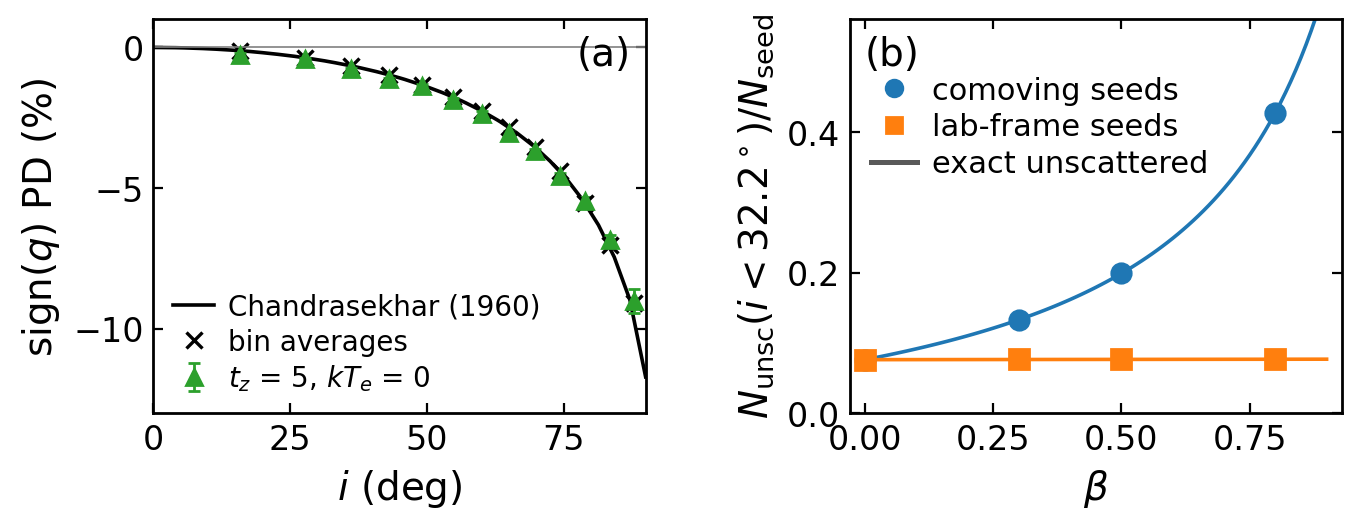}
  \caption{Results from comparing {\tt polrad-3D} with Chandrasekhar for an optically thick slab (left panel), and with analytical results for a thin slab. See text for details.}
  \label{f:cv}
\end{figure*}
We verified the {\tt polrad-3D} code by testing it against Chandrasekhar's classical results for optically thick stationary slabs \citep{1960ratr.book.....C}, against analytical results, and against the results published by \citep{2023ApJ...949L..10P}.
The left panel of Figure \ref{f:cv} compares the  {\tt polrad-3D} $q$ results for $t_z\,=\,5$ (green points) against Chandrasekhar's results for an indefinitely deep scattering atmosphere. We injected photons at 0.3 keV at the bottom of the plane parallel plasma slab ($h\,=\,0$), emitted with Chandrasekhar limb darkening and polarization. The simulations give the right Chandrasekhar results for all inclinations.
The right panel shows results from checking how our code simulates a moving emitting plasma layer. We tested this by simulating a slab with $t_z = 0.02$ of electrons at $k_{\rm B}T_{\rm e} = 100$\,keV moving with $\beta = 0$, 0.3, 0.5, and 0.8 upwards. Seed photons are injected isotropically in stationary frame (orange line) or in the plasma frame (blue line).
Photons generated at a height fraction $\zeta$ at a angle cosine of $\mu$ from the $z$-axis traverse an optical depth $t_z (1-\beta\mu)(1-\zeta)/\mu$ to the top with $1-\beta\mu$ accounting for the motion of the electrons or pairs. The fraction of unscattered photons leaving at $<32.2^{\circ}$ inclinations ($\mu > \mu_{\rm c} = 0.846$) is:
\begin{equation}
\frac{N_{\rm unsc}(i<32.2^{\circ})}{N_{\rm seed}}
  = \int_{\mu_{\rm c}}^{1} g(\mu)\,\frac{1-e^{-a(\mu)}}{a(\mu)}\,{\rm d}\mu ,
  \qquad a(\mu) = \frac{t_z\,(1-\beta\mu)}{\mu},
\end{equation}
with $g(\mu)$ being the lab-frame angular distribution of the seeds: $g = 1/2$ for emission isotropic in the lab frame, and $g = \frac{1}{2}(1-\beta^2)/(1-\beta\mu)^2$ for emission isotropic in the frame moving at
$\beta$. The right panel of Figure \ref{f:cv} shows that the code results match the analytical expression.

\begin{figure*}
  \centering
  \includegraphics[width=0.65\textwidth]{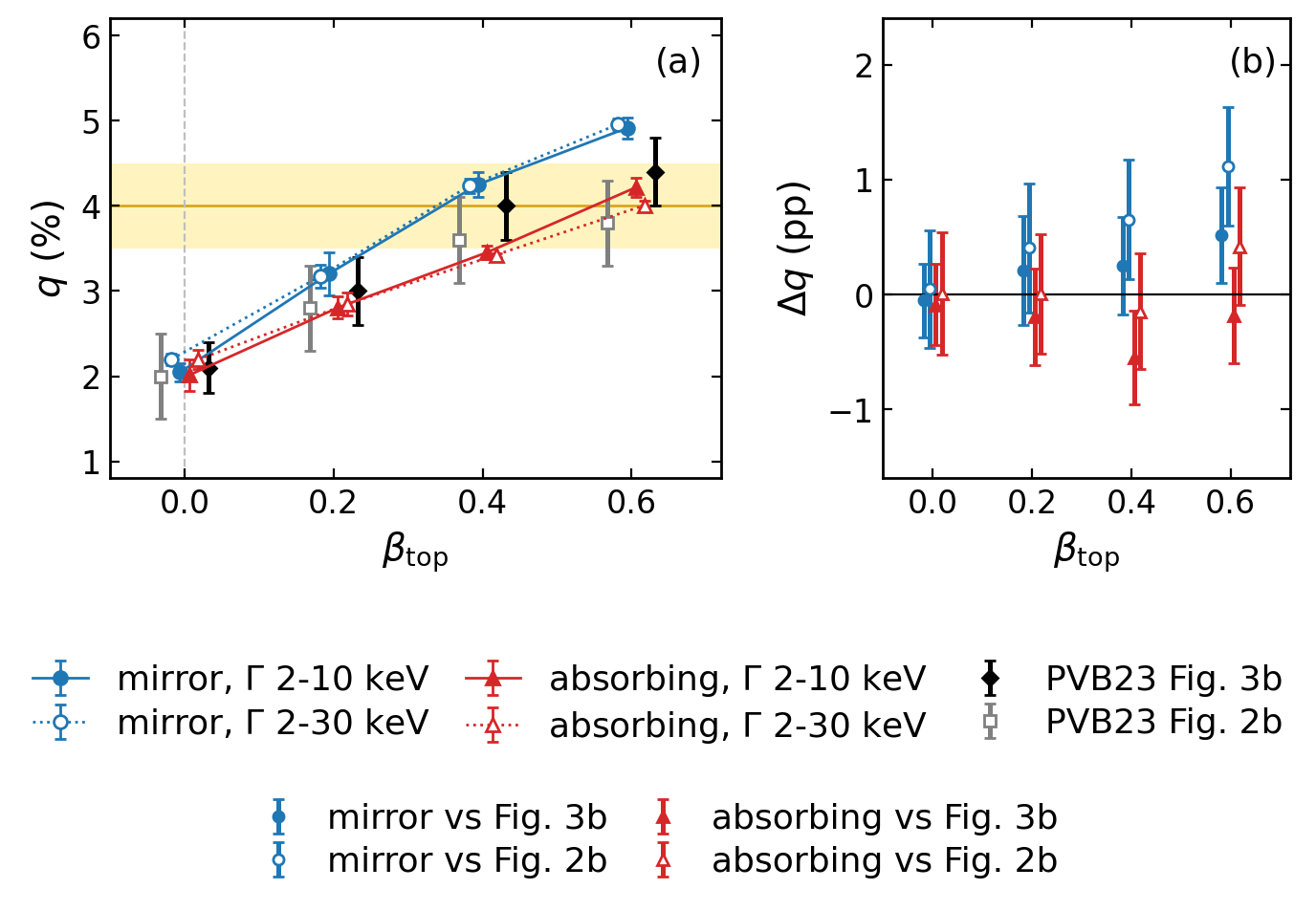}
  \caption{Comparison of results from {\tt polrad-3D} and the results from \citep{2023ApJ...949L..10P}. Left: comparison of absolute values. Right: deviation in percentage points. 
  All models assume plane parallel coronas ($h\,=\,0$). See text for details.
  }
  \label{f:pvb}
\end{figure*}
Figure \ref{f:pvb} compares the results from Figures 2 and 3 of \citep[][PVB23]{2023ApJ...949L..10P} against our results.
The red data points show our results with an absorbing lower corona boundary; the blue results show the same when we switch our lower absorbing layer to a reflecting layer to mimic the two-sided flow of PVB23.
Small differences include the fact that we limit the velocity to $\beta=\beta_0$ in the thin launching layer, and that we use 10 eV rather than 1 eV seed photons. The agreement is satisfactory although our $q$-values are slightly higher, especially when accounting for the fact that our inclination bin is centered at $27.8^{\circ}$ while their results 
are given at $30^{\circ}$. 
\label{app:test}
\end{document}